# Quantitatively Enhanced Reliability and Uniformity of High-κ Dielectrics on Graphene Enabled by Self-Assembled Seeding Layers

Vinod K. Sangwan[1], Deep Jariwala[1], Stephen A. Filippone[1,#], Hunter J. Karmel[1], James E. Johns[1], Justice M. P. Alaboson[1], Tobin J. Marks[1,2]\*, Lincoln J. Lauhon[1]\*, Mark C. Hersam[1,2,3]\*

[1]Department of Materials Science and Engineering, Northwestern University, Evanston, Illinois 60208

[2]Department of Chemistry, Northwestern University, Evanston, Illinois 60208

[3]Department of Medicine, Northwestern University, Evanston, Illinois 60208

ABSTRACT: The full potential of graphene in integrated circuits can only be realized with a reliable ultra-thin high-κ top-gate dielectric. Here, we report the first statistical analysis of the breakdown characteristics of dielectrics on graphene, which allows the simultaneous optimization of gate capacitance and the key parameters that describe large-area uniformity and dielectric strength. In particular, vertically heterogeneous and laterally homogenous $Al_2O_3$ and $HfO_2$ stacks grown *via* atomic-layer deposition and seeded by a molecularly thin perylene-3,4,9,10-tetracarboxylic dianhydride organic monolayer exhibit high uniformities (Weibull shape parameter β > 25) and large breakdown strengths (Weibull scale parameter, $E_{BD}$ > 7 MV/cm) that are comparable to control dielectrics grown on Si substrates.

The success of Si in integrated circuits (ICs) can be attributed to the fortuitous combination of several desirable properties exhibited by the interface between crystalline Si and its native amorphous oxide. For example, the technical ease of integration, low interfacial trap charge density, high interfacial barrier height with Si, and high breakdown reliability of silicon dioxide have contributed to the dominance of Si in ICs despite the lower charge carrier mobility of Si compared to competing semiconductors such as Ge and GaAs.[1,2] Today, an emerging electronic material, graphene, is being seriously considered for radio frequency (RF) electronics due to its high carrier mobility (~200,000 $cm^2$/Vs), high saturation velocity, and two-dimensional structure.[3-7] With these attributes, most of the applied research on top-gated graphene electronics has focused on achieving a high cut-off frequency, a high field-effect mobility, and current saturation.[8-20] However, in light of the aforementioned historical precedent set by Si, a reliable high-capacitance top-gate dielectric with wafer-scale uniformity will also be required for graphene to achieve the required scaling for high-frequency operation and seamless integration with existing technologies.[3,12,13,21] Towards this end, we report here the first extensive and systematic study of the breakdown characteristics of top-gated dielectrics deposited on graphene, thereby enabling the identification of optimized dielectric growth conditions and structures for high reliability and large-area uniformity.

Atomic layer deposition (ALD) has emerged as a ubiquitous technology for dielectric deposition in the microelectronics industry due to several compelling advantages such as large-area pin-hole free conformal coating, atomic-level thickness control, low-temperature growth, and compatibility with a wide range of materials.[22] However, the chemically inert nature of the graphene surface coupled with its hydrophobicity results in non-conformal ALD growth.[23, 24] Surface treatments of graphene to seed ALD *via* chemical functionalization with nitrogen dioxide,[25] oxidized metals,[26-28] or polymer coatings[11, 29, 30] can degrade the properties of the underlying graphene or reduce the overall gate capacitance. While physical vapor deposition[31] and oxidation of metal films on graphene can yield promising gate-dielectrics, these approach lack atomic level thickness control and suffer from the inherently rougher surfaces of evaporated metal films.[32, 33] Furthermore, efforts to seed ALD by ozone exposure demonstrated minimal perturbation to graphene at low growth temperatures but increased damage at higher temperatures.[34-36] Recently, ALD oxide growth has been seeded on graphene by self-assembled molecular layers of perylene-3,4,9,10-tetracarboxylic-3,4,9,10-dianhydride (PTCDA), which circumvents these issues and results in low-leakage, high-capacitance gate-dielectrics.[37-39] Here, we further demonstrate that PTCDA-seeded ALD growth provides a reliable pathway to vertically heterogeneous oxide stacks that simultaneously maximize gate capacitance, dielectric strength, and large-area homogeneity. In this manner, this work addresses the critical issues that underlie dielectric reliability and uniformity, thereby providing a pathway to wafer-scale graphene-based nanoelectronic circuits.

Dielectrics may break down due to intrinsic effects (e.g., Joule heating, impact ionization) or by extrinsic processing flaws (e.g., pin-holes, defects).[40] Notwithstanding the underlying mechanism, dielectric breakdown is characterized by a "weakest-link failure" event where a single short-circuit can trigger thermal runaway, leading to irreversible damage.[41, 42] Thus, dielectric breakdown is a highly stochastic process, requiring statistical analysis of a large number of test structures (e.g., capacitors) to understand the underlying breakdown mechanisms and predict reliability. Weakest-link failure events are typically characterized by the Weibull distribution.[43] Industry standards for predicting the lifetimes of electrical circuits are based on extreme value statistical analysis of time-dependent gate oxide breakdown (TDDB) measurements. Note that the voltage to breakdown ($V_{BD}$) distribution also follows a Weibull form for a constant voltage ramp-rate, which has been shown to be a robust predictor for TDDB.[44, 45] Therefore, for this first dielectric breakdown study on graphene, we conduct statistical analysis based on the $V_{BD}$ distribution. The 2-parameter Weibull cumulative distribution function (CDF) is given by:

$$F_{BD} = 1 - \exp\left[-\left(\frac{V_{BD}}{\alpha}\right)^{\beta}\right]$$

where α (the scale parameter) represents the voltage at which 63% of the devices fail, and β (the shape parameter) characterizes the width of the distribution. Consequently, a large value of α signifies high electrical breakdown strength, while a large value of β represents high uniformity. The Weibull CDF can be rearranged as:

$$\ln(-\ln(1 - F_{BD})) = \beta \ln(V_{BD}) - \beta \ln \alpha$$

which allows the Weibull parameters, α and β, to be extracted from the intercept and slope of a linear plot of $\ln(-\ln(1-F_{BD}))$ versus $\ln(V_{BD})$ (i.e., the Weibull plot). In terms of reliability, β <

1 signifies "infant-mortality" and the rate of failure decreases with increasing voltage (or time).[43] Such failures can be attributed to systematic defects caused by intrinsic flaws in the dielectric processing, and they result in "dead-on-arrival" products that are highly undesirable. Therefore, a β > 1 is one of the most important metrics for reliability where the failure rate increases with voltage (or time) due to common "wear-out."

In this study, capacitors were fabricated on epitaxially grown graphene (mixture of single-layer and bilayer graphene, Supporting S1) on 9 mm x 4.5 mm n-type 4H-SiC (EG-SiC) substrates. Less than two monolayers (< 2 MLs) of PTCDA were grown in a thermal evaporator (Fig. 1a, Supporting S1), followed by ALD growth of $Al_2O_3$ and $HfO_2$ dielectrics (Supporting S1). Capacitors of three different areas (area-1 = 20 μm x 20 μm, area-2 = 50 μm x 50 μm, and area-3 = 80 μm x 80 μm, see Figs. 1b, d) were fabricated by evaporating 100 nm thick Au through shadow masks. A dielectric thickness of 10 nm was chosen based on well-behaved current-voltage (I-V) characteristics. Thinner dielectrics (8 nm and 5 nm) exhibited significantly larger pin-hole density as discussed in Supporting S4. To simultaneously optimize capacitance, large-area uniformity, and dielectric strength, the two most commonly studied high-κ oxides, $Al_2O_3$ and $HfO_2$, were introduced in four different sample layering configurations. Fig. 1c shows the vertical structure of these four dielectric stacks: sample-1 (8 nm $HfO_2$ + 2 nm $Al_2O_3$ + PTCDA +EG-SiC); sample-2 (10 nm $Al_2O_3$ + PTCDA + EG-SiC); sample-3 (10 nm $HfO_2$ + PTCDA + EG-SiC); sample-4 (8 nm $Al_2O_3$ + PTCDA + EG-SiC). Capacitance-voltage (C-V) measurements on these samples (Fig. 1e) show the expected broad V-shape due to the quantum capacitance ($C_Q$) of graphene.[46, 47] The larger capacitance variation between devices in sample-3 is consistent with increased dielectric thickness variations from AFM images (Supporting S5). Flattening of the C-V plot on the left side (V < 0 V) can be attributed to the formation of a depletion region in the n-doped SiC. Therefore, a $C_Q$ model was fit to the right side of the curves to extract the capacitance of these dielectric stacks (Supporting S2). It should be noted that the $C_Q$ model does not fully account for all experimental conditions including the presence of a mixture of single-layer and bilayer graphene, gate-induced modulation of the depletion region in the SiC substrate, and the frequency-dependent capacitance resulting from series resistance at the ground contact (Supporting S2). However, these secondary effects are minimized by conducting the measurements at low frequency (1 kHz) and fitting the $C_Q$ model only for $V_g > 0$ V.

The capacitance of the four samples (sample-1 = 785 nF/cm$^2$, sample-2 = 545 nF/cm$^2$, sample-3 = 890 nF/cm$^2$, and sample-4 = 610 nF/cm$^2$) implies dielectric constants of the constituent oxides Al$_2$O$_3$ ($\kappa_{Al2O3}$ = 6.5) and HfO$_2$ ($\kappa_{HfO2}$ = 13) that are consistent with literature precedent[39] (Supporting S2). The extracted dielectric constants of the oxides are also consistent with the C-V response of the same dielectric stack deposited on Si control substrates (control-1: 8 nm HfO$_2$ + 2 nm Al$_2$O$_3$ + 1.8 nm native oxide + Si, Fig. 2a), see Supporting S2. A parallel-plate capacitor model also yields a PTCDA thickness (assuming $\kappa_{PTCDA}$ = 1.9, perpendicular to the molecular plane) of less than 0.5 nm in all the samples (Supporting Table S1). The extracted sub-2 ML thickness of PTCDA (1 ML thickness = 0.3 nm) from the C-V analysis is consistent with the absence of fluorescence in the Raman spectra for < 2 MLs PTCDA (Supporting S1).[37]

Capacitor I-V measurements were conducted in vacuum (< 10$^{-4}$ mbar) by biasing the Au electrode from 0-10 V at a ramp rate of 0.04 V/s. A positive voltage was chosen to minimize band-bending at the interface of the underlying substrates (n-SiC and n-Si). All samples showed leakage currents less than 10$^{-8}$ A/cm$^2$ for biases < 2 V (Supporting S3). At larger biases, the current begins to increase reversibly in the soft-breakdown regime (soft-breakdown mechanisms such as Fowler-Nordheim tunneling and Poole-Frenkel emission are discussed in Supporting S3).

Eventually, the voltage becomes high enough that the current spikes by 2-3 orders of magnitude and reaches the compliance limit (10 µA) within 0.01 V, allowing for a clear definition of the breakdown voltage ($V_{BD}$). This catastrophic event is characteristic of irreversible damage to the dielectric (i.e., hard breakdown). Fig. 2b shows the current density (J) plotted versus electric field (E) for sample-1 and control-1. The solid J-E curve corresponds to a capacitor with $V_{BD}$ in the middle of the Weibull distribution, while the dashed curves represent the two extreme values of $V_{BD}$, excluding low and high voltage tails. Thus, all of the capacitors measured for sample-1 and control-1 that follow the Weibull distribution show J-E curves that fall between the two corresponding dashed curves (I-V characteristics of all 210 sample-1 capacitors are shown in Supporting S3).

We first discuss the J-E curves in the soft-breakdown regime before proceeding to the statistical analysis of hard-breakdown characteristics. Both sample-1 and control-1 show indiscernible spreads in current density in Fig. 2b over the soft-breakdown regime (< 6 MV/cm) and a larger spread in the effective breakdown fields $E_{BD}$ ($E_{BD} = (V_{BD} - V_{Interface})/d$, d = total dielectric thickness, $V_{Interface}$ = voltage drop across graphene-SiC interface). Sample-1 capacitors break down earlier and have a wider spread in $E_{BD}$ than the control-1 capacitors. The current density in sample-1 is approximately an order of magnitude lower than that in control-1. This lower leakage current in sample-1 arises from its unique band-offset. In particular, lower leakage currents in high-κ (small band-gap) dielectrics are commonly achieved by creating a large band-offset *via* incorporation of an ultra-thin interfacial layer of a low-κ (large band-gap) material.[48] However, this mechanism fails to explain the characteristics of the present devices because the interfacial layer of $SiO_2$ in control-1 has a larger band-offset than PTCDA or $Al_2O_3$ in sample-1. Consequently, we attribute the lower current density in sample-1 to a Schottky barrier at the interface between the epitaxial graphene and the n-type 4H-SiC.[49] Previously, C-V measurements on EG-SiC have shown Fermi-level ($E_F$) pinning of graphene approximately 0.49 eV above the charge-neutrality point.[49] Quantum capacitance model fits of C-V curves (Fig. 1e) also yield a charge density (n) of ~5 x $10^{12}$/cm$^2$ (Supporting Table S1), resulting in a Fermi energy offset of 0.28 eV, calculated from $E_F = \hbar v_F \sqrt{\pi n}$, where $\hbar$ is Planck's constant divided by $2\pi$ and $v_F$ (~$10^8$ cm/s) is the Fermi velocity of graphene. The discrepancy in the derived Fermi level could be due to bilayer graphene regions in EG-SiC and/or additional doping introduced by the dielectric growth process. In either case, a Schottky barrier in addition to a triangular potential barrier in the dielectric can explain the reduced quantum mechanical tunneling probability and thus the lower current density. Note that this Schottky contact also reduces the overall voltage across the dielectrics by 0.28 V, which is then subtracted from the applied voltage to obtain the effective electric field (E) in the dielectrics for all samples except control-2 (consisting of same oxide stack as sample-1 (8 nm $HfO_2$ + 2 nm $Al_2O_3$) grown directly on EG-SiC without PTCDA, Fig. 3c) where 20% of the devices break at $V_{BD}$ < 0.28 V due to excessive pin-holes (Supporting S5).

A Weibull plot of $V_{BD}$ distributions for sample-1 capacitors of different areas is shown in Fig. 3a (see $V_{BD}$ histograms in Supporting S3). A good linear fit ($r^2$ > 0.95) to the majority of the data (>85%, excluding low voltage tails that are visibly separated from the majority of the data) indicates that the $V_{BD}$ data are described well by a 2-parameter Weibull distribution. Weibull parameters and 95% confidence bounds were extracted using a previously reported methodology (Supporting Table S2).[50, 51] Since Weibull fits assume a random distribution of breakdown locations within the dielectric, we confirm the validity of this assumption explicitly by scaling all capacitors to a common area of 1 mm$^2$. For randomly distributed defects, $V_{BD}$ data is expected to obey the following area (A) scaling law:[42]

$$\ln(-\ln(1-F_1)) - \ln(-\ln(1-F_2)) = \ln(A_1/A_2).$$

Therefore, random defects should result in statistically identical β values for different areas, whereas α should scale as $\alpha_1/\alpha_2 = [A_2/A_1]^{1/\beta}$.[42] Since shadow masking results in up to 5% variation in the capacitor areas, the average area of each sample was directly confirmed by optical contrast imaging (Supporting S4 and Table S2). A Weibull plot of all 210 area-scaled sample-1 data also fits well to a straight line (r$^2$ > 0.9) as shown in Fig. 3b. However, area-1 capacitors show a larger β (103) than area-2 (27.3) and area-3 (29) devices, possibly due to the spatial distribution of breakdown locations having a characteristic length scale. Also note that the $V_{BD}$ distribution of the area-1 capacitors has multiple values of $V_{BD}$ that are identical within experimental error (within 0.01 V), which results in an artifact of infinite slope and thus an overestimated β.

To quantify the positive effect of PTCDA seeding on reliability and dielectric strength, we also consider a control sample (control-2) consisting of the same oxide stack as sample-1 (8 nm HfO$_2$ + 2 nm Al$_2$O$_3$) grown directly on EG-SiC without PTCDA (Fig. 3c). The area-scaled Weibull plot (Fig. 3d) of $V_{BD}$ for control-2 shows that the $V_{BD}$ distribution does not scale with area. The extracted β from the majority of the data (>90%, excluding low-voltage tails) is 0.96 for area-1 and 0.98 for area-3, although β = 1 falls within the 95% confidence bounds (Supporting Table S2). Note that 10% of the control-2 capacitors still break down at an electric field greater than 4 MV/cm. Therefore, even though direct ALD growth on graphene can produce isolated operational devices, a seeding layer is needed to achieve $\beta > 1$ (i.e., large-area uniformity) and large breakdown fields.

We also explored how PTCDA-seeded ALD dielectrics on graphene compare with ALD oxides on Si substrates. To address this issue, we conducted a breakdown study of capacitors (Fig. 2a) on Si substrates (control-1). Control-1 $V_{BD}$ data follow a Weibull distribution and scale well with area (Supporting S4). Area-scaled $E_{BD}$ distributions of all samples (sample-1, -2, -3) and controls (control-1, -2) are displayed together in the Weibull plot of Fig. 4a. The atomically thin PTCDA seeding layer shifts the $E_{BD}$ distributions of all three samples from control-2 close to control-1; i.e., the PTCDA seeding increases the dielectric strength by more than an order of magnitude ($\alpha(E_{BD}) = 0.45$ MV/cm for control-2, whereas $\alpha(E_{BD}) = 6.83 - 7.31$ MV/cm for the three PTCDA-seeded samples) and remains only 15% lower than control-1 ($\alpha(E_{BD}) = 8.17$ MV/cm). In addition, PTCDA contributes to a significant increase in large-area dielectric uniformity ($\beta > 25$ for sample-1, $\beta \leq 1$ for control-2).

For an overall assessment of dielectric quality, uniformity and dielectric strength should be analyzed concurrently with capacitance. Dielectric capacitance (C) and dielectric strength ($V_{BD}$) determine the largest reversible modulation of carrier concentration (n) in the semiconductor channel ($n = C(V_{BD} - V_{Interface})/e = \kappa\varepsilon_0 E_{BD}/e$; e = electronic charge, $\varepsilon_0$ = vacuum permittivity). Thus, the intrinsic dielectric property is the maximum displacement field (scale parameter $D_{max} = \kappa\varepsilon_0\alpha(E_{BD})$). Note that although capacitance can also be increased by thinning the dielectric, surface roughness and defects in the EG-SiC wafers limited the thickness to approximately 10 nm (Supporting S3), which is significantly thicker than the fundamental limits of quantum mechanical tunneling (~1 nm).

We now discuss the role of vertical heterogeneity in the oxide stack in relation to enhancing the dielectric performance metrics. Fig. 4b shows zoomed-in $E_{BD}$ distributions of area-3 capacitors on samples-1, -2, and -3 from Fig. 4a (see Supporting S4 for area-scaled Weibull plots of sample-2 and -3). Uniformity of sample-1 (95% confidence bounds of β: 23.8 – 35.3) is higher than both sample-2 (β: 17.0 – 25.5) and sample-3 (β: 9.5 – 16.5). Although the effective dielectric strength scale parameter (α($E_{BD}$)) of sample-1 is only marginally greater than those of sample-2 and sample-3, the occurrences of "leaky capacitors" as soon as measurements were begun is significantly smaller in sample-1 (<1%) than sample-2 (>10%) and sample-3 (>17%) (Supporting S3 and Table S2). In addition, the sample-3 dielectric was found to be mechanically fragile and showed a tendency to peel off during probing. Therefore, we conclude that PTCDA seeds ALD growth of $Al_2O_3$ more effectively than $HfO_2$ on graphene, which was also suggested by microscopy analysis in a previous study.[39] Owing to the 2x greater κ value for $HfO_2$ versus $Al_2O_3$, sample-1 also shows a larger value of $D_{max}$ (6.02 µC/cm$^2$) than sample-2 (3.32 µC/cm$^2$), see Supporting Table S1. Note that the Schottky contact between the SiC substrate and the ground probe may introduce a uniform shift in the scale parameters of the samples, although the shape parameters are unaffected by a constant shift in $V_{BD}$.

Empirically, the dielectric constant and breakdown fields of oxide dielectrics are known to exhibit an inverse relationship.[52] From that perspective, we highlight two surprising observations. First, in spite of higher effective κ, sample-1 shows higher uniformity (β), larger dielectric strength (α($E_{BD}$)), and a significantly reduced fraction of "leaky capacitors" compared to sample-2. This observation suggests that a sequential ALD growth of two different oxides likely interferes with pin-hole formation, perhaps through misalignment of randomly distributed pin-holes in each of the two different layers. Therefore, the attractive attributes of sample-1 may originate not only from advantageously tailored chemistry but also from the physical aspects of vertical heterogeneity. Second, in spite of the 13% smaller capacitance of sample-1 versus sample-3, $D_{max}$ of sample-1 is only 6% lower than that of sample-3. The greater $\alpha(E_{BD})$ and $D_{max}$ values of sample-1 can be attributed to the interfacial $Al_2O_3$ layer in sample-1 that has a larger band-gap (8.8 eV), larger band-offset, and larger breakdown strength than $HfO_2$ in sample-3.[52] These results are in agreement with experimental[53] as well as theoretical[54] reports that have demonstrated improved characteristics in Si devices that utilize $Al_2O_3$ + $HfO_2$ oxide stacks in place of $HfO_2$ alone. The improved uniformity and breakdown strength of sample-1 is also consistent with surface morphology analysis of the samples *via* atomic force microscopy (Supporting S5).

In conclusion, we have addressed large-area reliability and breakdown limits of gate dielectrics grown on graphene. A molecularly thin organic seeding layer, PTCDA, leads to significant improvements in dielectric performance limits (e.g., increasing the Weibull scale parameter and shape parameter in the $E_{BD}$ distribution by more than an order of magnitude) for ALD-grown oxides on graphene. Furthermore, we find that vertically heterogeneous oxide stacks provide additional advantages from the perspectives of large-area homogeneity and maximum displacement field. In particular, a 2 nm thick interfacial $Al_2O_3$ layer is shown to enable robust ($\beta > 1$) growth of high-$\kappa$ $HfO_2$, thus achieving a high breakdown field of >7 MV/cm at the highest capacitance (785 nF/cm$^2$) thus far reported on large area epitaxial graphene substrates. These results, coupled with the ability of PTCDA-seeded ALD top dielectrics to yield competitive graphene field-effect transistors (see Supporting S6), present clear future opportunities for graphene-based large-scale integrated circuits.

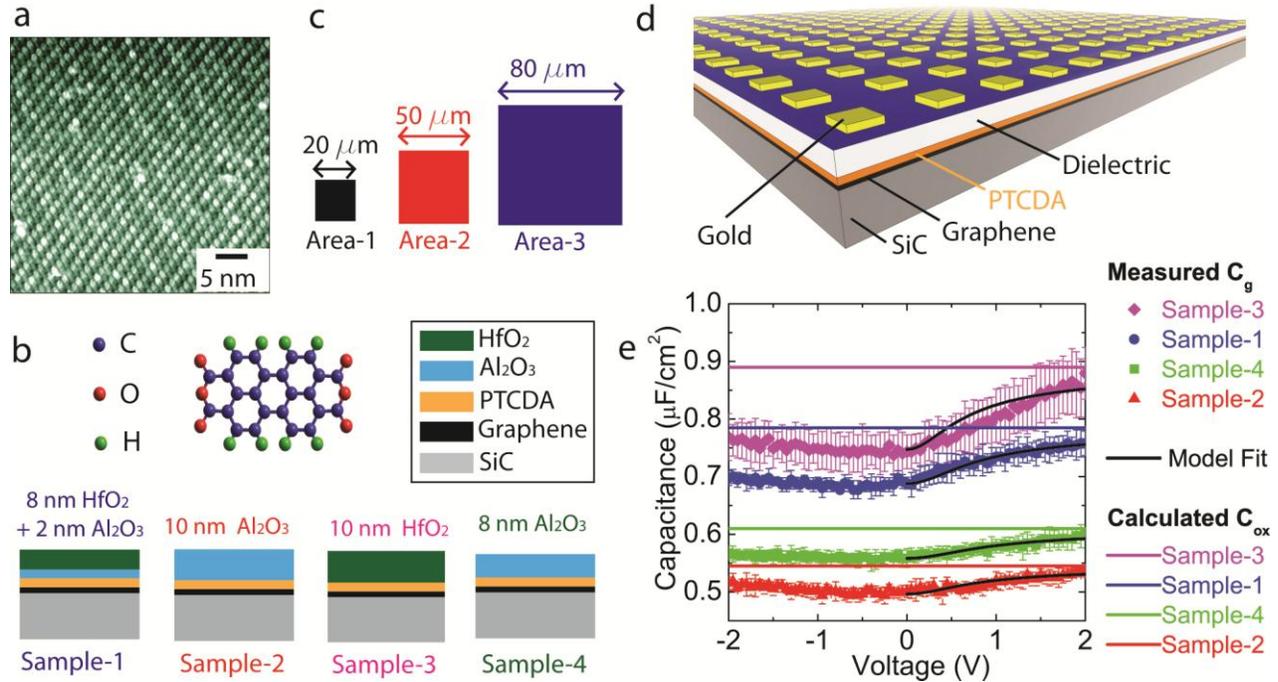

**Figure 1.** (a) High-resolution scanning tunneling microscopy (STM) image of a PTCDA monolayer on epitaxial graphene on SiC. (b) PTCDA molecular structure and the four different sample structures consisting of different combinations of $Al_2O_3$ and $HfO_2$ layers. Color codes for the materials (black box) and samples are retained throughout all figures. (c) Square capacitors of three different areas (area-1: 20 μm x 20 μm, area-2: 50 μm x 50 μm, and area-3: 80 μm x 80 μm) fabricated on the each sample. (d) Schematic of an array of capacitors on a large area dielectric. (e) Capacitance-voltage curves of sample-1, sample-2, sample-3, and sample-4 at a frequency of 1 kHz. Data points are the average of 5 capacitors, and the error bars represent standard deviations. The black lines are fit to the quantum capacitance model (Supporting S2). The solid horizontal lines represent the extracted capacitances of the oxide stacks in each of the samples.

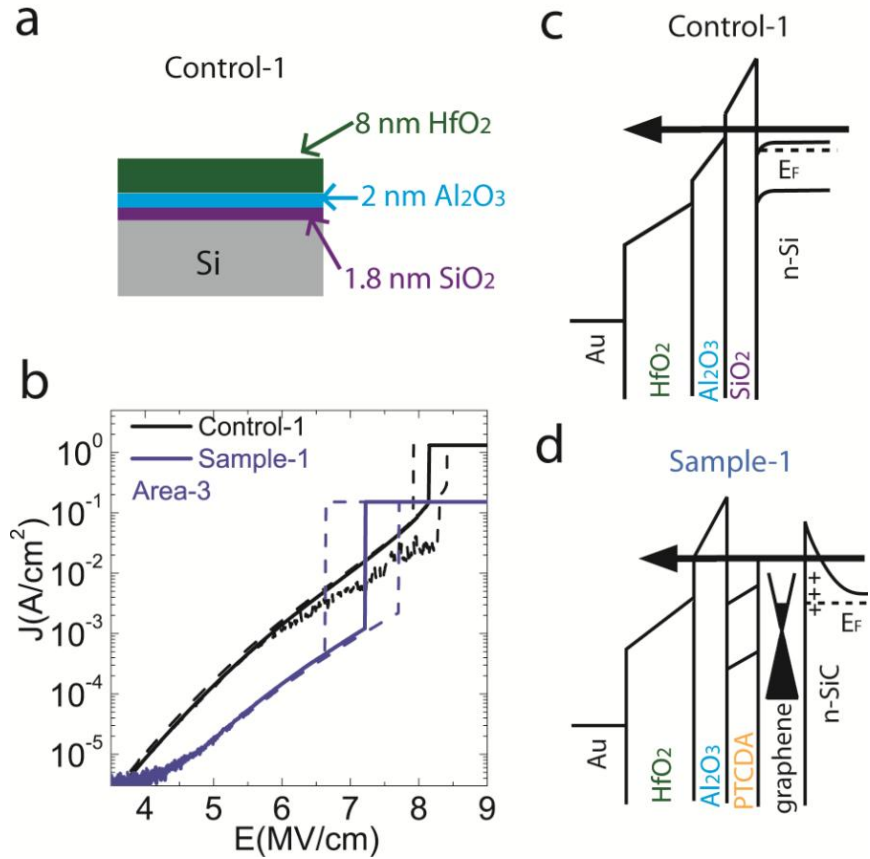

**Figure 2.** (a) Schematic of control-1 capacitors on an n-doped Si substrate. (b) Current density versus electric field (J-E) characteristics of sample-1 and control-1 capacitors. The solid and the dashed lines correspond to the capacitor in the middle and capacitors on the extremes of the Weibull distribution, respectively. (c) Schematic band-diagram of a control-1 capacitor when the Au electrode is biased to a positive voltage. (d) Schematic band-diagram of a sample-1 capacitor. Note that n-doped graphene on SiC forms a Schottky junction, which presents an additional tunnel barrier for the leakage current.

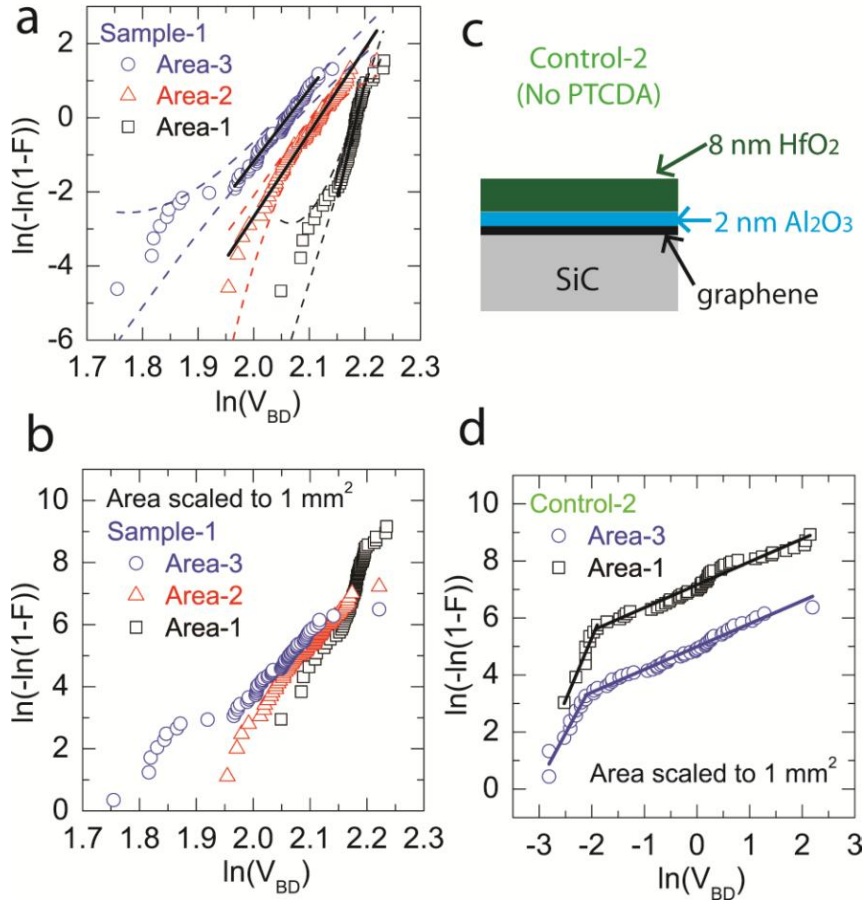

**Figure 3.** (a) Weibull plot of the breakdown voltage ($V_{BD}$) distribution for sample-1 capacitors. The number of data points for area-1, area-2, and area-3 are 74, 66, and 70, respectively. The dashed lines represent 95% confidence bounds. Linear fits (black lines) are performed on the majority of the data (excluding low-voltage tails) to extract Weibull parameters. (b) Weibull plot of the sample-1 $V_{BD}$ data after scaling to a common area of 1 mm$^2$. (c) Schematic of control-2 capacitors fabricated on EG-SiC without PTCDA. (d) Weibull plot of control-2 $V_{BD}$ data after scaling to a common area of 1 mm$^2$. Linear fits (solid lines) are performed on the majority of the data (excluding low-voltage tails) to extract Weibull parameters.

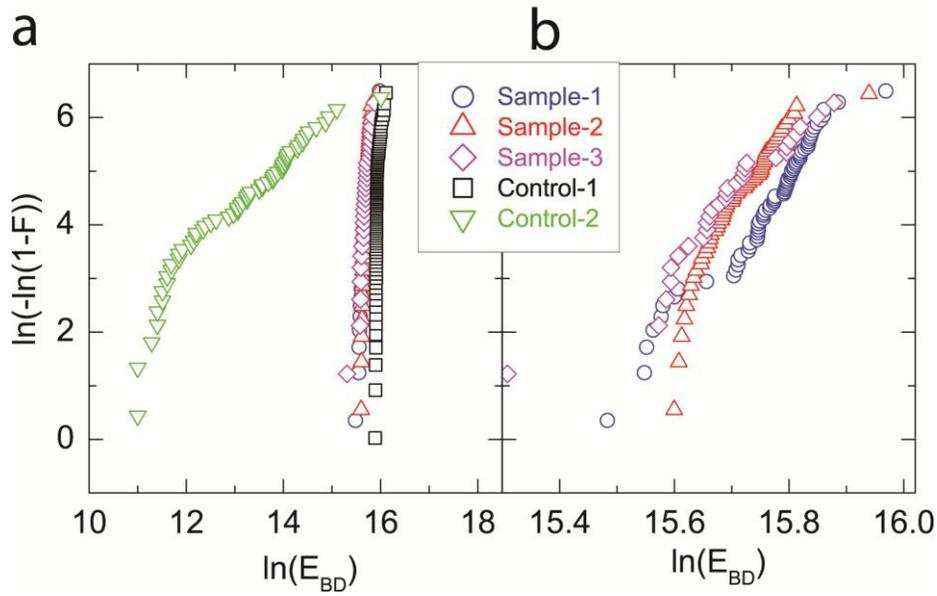

**Figure 4.** (a) Weibull plot of the breakdown fields $E_{BD}$ (V/cm) for the three samples and two controls. $E_{BD}$ data of area-3 capacitors on each sample were scaled to a common area of 1 mm$^2$. (b) The same Weibull plot is displayed over a narrower $E_{BD}$ range to compare the $E_{BD}$ distribution for sample-1, sample-2, and sample-3.

ASSOCIATED CONTENT

**Supporting Information**. Detailed description of graphitization of SiC; deposition of PTCDA; Raman analysis; atomic layer deposition of dielectrics; quantum capacitance modeling and capacitance-voltage analysis; dielectric breakdown measurements and reliability analysis of all samples and controls; two tables of Weibull parameters; atomic force microscopy analysis; top-gated graphene field-effect transistors. This material is available free of charge via the Internet at http://pubs.acs.org.

AUTHOR INFORMATION

**Corresponding Authors**

*Emails: t-marks@northwestern.edu; lauhon@northwestern.edu; m-hersam@northwestern.edu

**Present Addresses**

# Department of Materials Science and Engineering, Johns Hopkins University, Baltimore, MD, 21218

**Author Contributions**

The manuscript was written through contributions of all authors. All authors have given approval to the final version of the manuscript.

**Notes**

The authors declare no competing financial interest.

ACKNOWLEDGMENT: This research was supported by the Office of Naval Research (N00014-11-1-0463), a W. M. Keck Foundation Science and Engineering Grant, and the Materials Research Science and Engineering Center (MRSEC) of Northwestern University (NSF DMR-1121262). The authors thank Dr. P. Prabhumirashi and K. Yoon for helpful discussions, and K. Stallings and B. Myers for assistance with evaporation and electron beam lithography, respectively. This research made use of the NUANCE Center at Northwestern University, which is supported by NSF-NSEC, NSF-MRSEC, Keck Foundation, and the State of Illinois.

REFERENCES

1. Sofield C. J.; Stoneham A. M. *Semicond. Sci. Technol.* **1995,** 10, 215.

2. Verwey, J. F.; Amerasekera, E. A.; Bisschop, J. *Rep. Prog. Phys.* **1990,** 53, 1297.

3. Schwierz, F. *Nat. Nanotech.* **2010,** 5, 487.

4. Novoselov, K. S.; Geim, A. K.; Morozov, S. V.; Jiang, D.; Zhang, Y.; Dubonos, S. V.; Grigorieva, I. V.; Firsov, A. A. *Science* **2004,** 306, 666.

5. Jariwala, D.; Sangwan, V. K.; Lauhon, L. J.; Marks, T. J.; Hersam, M. C. *Chem. Soc. Rev.* **2013,** DOI: 10.1039/c2cs35335k.

6. Novoselov, K. S.; Falko, V. I.; Colombo, L.; Gellert, P. R.; Schwab, M. G.; Kim, K. *Nature* **2012,** 490, 192.

7. Banerjee, S. K.; Register, L. F.; Tutuc, E.; Basu, D.; Seyoung, K.; Reddy, D.; MacDonald, A. H. *Proceed. IEEE* **2010,** 98, 2032.

8. Dharmendar, R.; Leonard, F. R.; Gary, D. C.; Sanjay, K. B. *J. Phys. D: Appl. Phys.* **2011,** 44, 313001.

9. Morozov, S. V.; Novoselov, K. S.; Katsnelson, M. I.; Schedin, F.; Elias, D. C.; Jaszczak, J. A.; Geim, A. K. *Phys. Rev. Lett.* **2008,** 100, 016602.

10. Dean, C. R.; Young, A. F.; Meric I.; Lee C.; Wang L.; Sorgenfrei S.; Watanabe K.; Taniguchi T.; Kim P.; Shepard, K. L.; Hone J. *Nat. Nanotech.* **2010,** 5, 722.

11. Farmer, D. B.; Chiu, H.-Y.; Lin, Y.-M.; Jenkins, K. A.; Xia, F.; Avouris, P. *Nano Lett.* **2009,** 9, 4474.

12. Lin, Y.-M.; Dimitrakopoulos, C.; Jenkins, K. A.; Farmer, D. B.; Chiu, H.-Y.; Grill, A.; Avouris, P. *Science* **2010**, 327, 662.

13. Lin, Y.-M.; Valdes-Garcia, A.; Han, S.-J.; Farmer, D. B.; Meric, I.; Sun, Y.; Wu, Y.; Dimitrakopoulos, C.; Grill, A.; Avouris, P.; Jenkins, K. A. *Science* **2011,** 332, 1294.

14. Wu, Y.; Lin, Y.-m.; Bol, A. A.; Jenkins, K. A.; Xia, F.; Farmer, D. B.; Zhu, Y.; Avouris, P. *Nature* **2011,** 472, 74.

15. Xu, H.; Zhang, Z.; Xu, H.; Wang, Z.; Wang, S.; Peng, L.-M. *ACS Nano* **2011,** 5, 5031.

16. Bolotin, K. I.; Sikes, K. J.; Jiang, Z.; Klima, M.; Fudenberg, G.; Hone, J.; Kim, P.; Stormer, H. L. *Solid State Commun.* **2008,** 146, 351.

17. Meric, I.; Han, M. Y.; Young, A. F.; Ozyilmaz, B.; Kim, P.; Shepard, K. L. *Nat. Nanotech.* **2008,** 3, 654.

18. Obeng, Y.; Srinivasan, P. *Electrochem. Soc. Interface* **2011,** 20, 47.

19. Liao, L.; Bai, J.; Qu, Y.; Lin, Y.-c.; Li, Y.; Huang, Y.; Duan, X. *Proc. Natl. Acad. Sci. U.S.A.* **2010,** 107, 6711.

20. Liao, L.; Lin, Y.-C.; Bao, M.; Cheng, R.; Bai, J.; Liu, Y.; Qu, Y.; Wang, K. L.; Huang, Y.; Duan, X. *Nature* **2010,** 467, 305.

21. Liao, L.; Duan, X. *Mater. Sci. Eng.: R: Reports* **2010,** 70, 354.

22. George, S. M. *Chem. Rev.* **2009,** 110, 111.

23. Xuan, Y.; Wu, Y. Q.; Shen, T.; Qi, M.; Capano, M. A.; Cooper, J. A.; Ye, P. D. *Appl. Phys. Lett.* **2008,** 92, 013101.

24. Garces, N. Y.; Wheeler, V. D.; Gaskill, D. K. *J. Vac. Sci. Technol., B* **2012,** 30, 030801.

25. Lin, Y.-M.; Jenkins, K. A.; Valdes-Garcia, A.; Small, J. P.; Farmer, D. B.; Avouris, P. *Nano Lett.* **2008,** 9, 422.

26. Shen, T.; Gu, J. J.; Xu, M.; Wu, Y. Q.; Bolen, M. L.; Capano, M. A.; Engel, L. W.; Ye, P. D. *Appl. Phys. Lett.* **2009,** 95, 172105.

27. Fallahazad, B.; Lee, K.; Lian, G.; Kim, S.; Corbet, C. M.; Ferrer, D. A.; Colombo, L.; Tutuc, E. *Appl. Phys. Lett.* **2012,** 100, 093112.

28. Robinson, J. A.; LaBella, M.; Trumbull, K. A.; Weng, X.; Cavelero, R.; Daniels, T.; Hughes, Z.; Hollander, M.; Fanton, M.; Snyder, D. *ACS Nano* **2010,** 4, 2667.

29. Meric, I.; Dean, C. R.; Young, A. F.; Baklitskaya, N.; Tremblay, N. J.; Nuckolls, C.; Kim, P.; Shepard, K. L. *Nano Lett.* **2011,** 11, 1093.

30. Shin, W. C.; Kim, T. Y.; Sul, O.; Cho, B. J. *Appl. Phys. Lett.* **2012,** 101, 033507.

31. Wu, Y. Q.; Ye, P. D.; Capano, M. A.; Xuan, Y.; Sui, Y.; Qi, M.; Cooper, J. A.; Shen, T.; Pandey, D.; Prakash, G.; Reifenberger, R. *Appl. Phys. Lett.* **2008,** 92, 092102.

32. Kim, S.; Nah, J.; Jo, I.; Shahrjerdi, D.; Colombo, L.; Yao, Z.; Tutuc, E.; Banerjee, S. K. *Appl. Phys. Lett.* **2009,** 94, 062107.

33. Xu, H.; Zhang, Z.; Wang, Z.; Wang, S.; Liang, X.; Peng, L.-M. *ACS Nano* **2011,** 5, 2340.

34. Lee, B.; Mordi, G.; Kim, M. J.; Chabal, Y. J.; Vogel, E. M.; Wallace, R. M.; Cho, K. J.; Colombo, L.; Kim, J. *Appl. Phys. Lett.* **2010,** 97, 043107.

35. Lee, B.; Park, S.-Y.; Kim, H.-C.; Cho, K.; Vogel, E. M.; Kim, M. J.; Wallace, R. M.; Kim, J. *Appl. Phys. Lett.* **2008,** 92, 203102.

36. Jandhyala, S.; Mordi, G.; Lee, B.; Lee, G.; Floresca, C.; Cha, P.-R.; Ahn, J.; Wallace, R. M.; Chabal, Y. J.; Kim, M. J.; Colombo, L.; Cho, K.; Kim, J. *ACS Nano* **2012,** 6, 2722.

37. Johns, J. E.; Karmel, H. J.; Alaboson, J. M. P.; Hersam, M. C. *J. Phys. Chem. Lett.* **2012,** 3, 1974.

38. Wang, Q. H.; Hersam, M. C. *Nat. Chem.* **2009,** 1, 206.

39. Alaboson, J. M. P.; Wang, Q. H.; Emery, J. D.; Lipson, A. L.; Bedzyk, M. J.; Elam, J. W.; Pellin, M. J.; Hersam, M. C. *ACS Nano* **2011,** 5, 5223.

40. Sah, C. T. *Solid-State Electron.* **1990,** 33, 147.

41. Le, J.-L.; Bazant, Z. P.; Bazant, M. Z. *J. Appl. Phys.* **2009,** 106, 104119.

42. Wu, E. Y.; Vollertsen, R. P. *IEEE Trans. Electron Dev.* **2002,** 49, 2131.

43. Weibull, W. *J. Appl. Mech.* **1951,** 18.

44. Dissado, L. A.; Fothergill, J. C.; Wolfe, S. V.; Hill, R. M. *IEEE Trans. Elec. Insul.* **1984,** EI-19, 227.

45. Hill, R. M.; Dissado, L. A. *J. Phys. C: Solid State Phys.* **1983,** 16, 2145.

46. Fang, T.; Konar, A.; Xing, H.; Jena, D. *Appl. Phys. Lett.* **2007,** 91, 092109.

47. Xia, J.; Chen, F.; Li, J.; Tao, N. *Nat. Nanotech.* **2009,** 4, 505.

48. Banerjee, S.; Shen, B.; Chen, I.; Bohlman, J.; Brown, G.; Doering, R. *J. Appl. Phys.* **1989,** 65, 1140.

49. Sonde, S.; Giannazzo, F.; Raineri, V.; Yakimova, R.; Huntzinger, J. R.; Tiberj, A.; Camassel, J. *Phys. Rev. B* **2009,** 80, 241406.

50. Schlitz, R. A.; Ha, Y.-g.; Marks, T. J.; Lauhon, L. J. *ACS Nano* **2012,** 6, 4452.

51. Schlitz, R. A.; Yoon, K.; Fredin, L. A.; Ha, Y.-g.; Ratner, M. A.; Marks, T. J.; Lauhon, L. J. *J. Phys. Chem. Lett.* **2010,** 1, 3292.

52. McPherson, J.; Kim, J. Y.; Shanware, A.; Mogul, H. *Appl. Phys. Lett.* **2003,** 82, 2121.

53. Son, J.-Y.; Jeong, S.-W.; Kim, K.-S.; Roh, Y. *J Korean Phys. Soc.* **2007,** 51.

54. Conde, A.; Martínez, C.; Jiménez, D.; Miranda, E.; Rafí, J. M.; Campabadal, F.; Suñé, J. *Solid-State Electron.* **2012,** 71, 48.

**Table of Contents Figure**

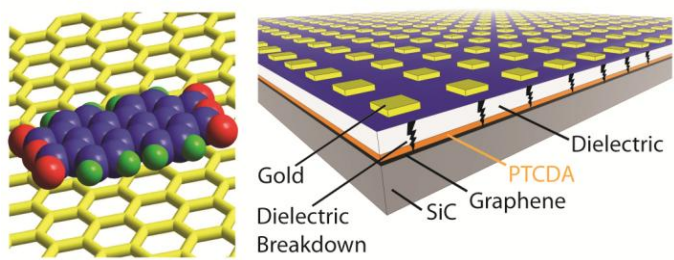